\documentclass[showkeys,showpacs,superscriptaddress,nofootinbib,prd]{revtex4-2}
\usepackage{amsmath}
\usepackage{amssymb}
\usepackage{graphicx}
\usepackage{yfonts}
\usepackage{ulem,xcolor}
\usepackage{color,soul}
\usepackage{amsthm}
\usepackage{yhmath}

\newtheorem{theorem}{Theorem}[section]
\theoremstyle{definition}
\newtheorem{definition}{Definition}[section]

\begin{document}
	
\title{Quantum measure as a necessary ingredient in quantum gravity and  modified gravities
	}
	
\author{
	Vladimir Dzhunushaliev
}
\email{v.dzhunushaliev@gmail.com}
	
\affiliation{
	Department of Theoretical and Nuclear Physics,  Al-Farabi Kazakh National University, Almaty 050040, Kazakhstan
}

\affiliation{
	Academician J.~Jeenbaev Institute of Physics of the NAS of the Kyrgyz Republic, 265 a, Chui Street, Bishkek 720071, Kyrgyzstan
}

\affiliation{
	International Laboratory for Theoretical Cosmology, Tomsk State University of Control Systems and Radioelectronics (TUSUR),
	Tomsk 634050, Russia
}

\author{Vladimir Folomeev}
\email{vfolomeev@mail.ru}

\affiliation{
	Academician J.~Jeenbaev Institute of Physics of the NAS of the Kyrgyz Republic, 265 a, Chui Street, Bishkek 720071, Kyrgyzstan
}

\affiliation{
	International Laboratory for Theoretical Cosmology, Tomsk State University of Control Systems and Radioelectronics (TUSUR),
	Tomsk 634050, Russia
}
	
\date{\today}
	
\begin{abstract}
We suggest commutation relations for a quantum measure. In one version of these relations, the right-hand side 
takes account of the presence of curvature of space; in the simplest case, this yields the action of general relativity. 
We consider the cases of the quantization of the measure on spaces of constant curvature and show that in this case the commutation 
relations for the quantum measure are analogues of commutation relations in loop quantum gravity. It is assumed that, 
in contrast to loop quantum gravity, a triangulation of space is a necessary trick for quantizing such a nonlocal  quantity 
like a measure; in doing so, the space remains a smooth manifold. We consider the self-consistent problem of the interaction of the 
quantum measure and classical gravitation. It is shown that this inevitably leads to the appearance of modified gravities. 
Also, we consider the problem of defining the Euler-Lagrange equations for a matter field in the background of a space 
endowed with quantum measure.
\end{abstract}
	
	
	
\date{\today}
	
\maketitle
	
\section{Introduction}

Modified gravities lead to various interesting consequences, among which are the Starobinsky inflation~\cite{Starobinsky:1980te}, the explanation of the present accelerated expansion of the Universe~\cite{Nojiri:2017ncd}, etc. But the following fundamental question arises as to how from all possible versions of modified gravities Nature chose the one? The truth is that there is a continuum of such versions, i.e., it is not even a countable set, and it is evident that if  Nature chose any variant of them, this means that it differs somehow from other ones. If one compares this situation with general relativity, the answer is very simple: general relativity is the simplest geometrical theory of gravitation 
with second-order field equations and with the Lagrangian proportional to the scalar curvature. The answer to the above question can be as follows: 
modified gravities are an approximate description of quantum gravity based on general relativity, and various versions of modified gravities are realized 
in different physical situations. In this paper, following this assumption, we show that the quantization of the measure leads to modified gravities 
whose form depends on the particular physical situation.

There are not so many investigations devoted to a measure as a physical quantity. The following studies in this direction may be noted. 
In Refs.~\cite{Guendelman:1996qy,Guendelman:1996jr,Guendelman:2012gg}, instead of a measure determined by a metric, one introduces the so-called
dynamical measure $
d \phi^1 \wedge d \phi^2 \wedge d \phi^3 \wedge d \phi^4
$ with the scalar fields $\phi^i, i = 1,2,3,4$, which are determined using some dynamical equations. In Ref.~\cite{Finster:2023rkv}, 
a comparison of modified measure theories (an approach to modified theories of gravity) with the theory of causal fermion systems
(an approach to unify quantum theory with general relativity) is carried out.
In Ref.~\cite{Gronwald:1997ei},  the options for defining a volume element $\sigma$, which can be used for physical theories and defined as a ``dilaton'' field, are discussed. 

Modified gravities are very fascinating for many reasons, and one of them is that they may serve as alternative models of dark energy. In particular, 
in Ref.~\cite{Nojiri:2003ft}, modified theory of gravity is suggested within which the terms with positive powers of the curvature support the inflationary 
epoch, while the terms with negative powers of the curvature serve as effective dark energy, supporting the current cosmic acceleration. 
In Ref.~\cite{Lidsey:2002zw}, braneworld cosmology for a domain wall embedded in the charged
anti-de Sitter-Schwarzschild black hole of the five-dimensional Einstein-Gauss-Bonnet-Maxwell theory is considered. In Ref.~\cite{Chakraborty:2020ifg}, 
the bound on the compactness of a stellar object in pure Lovelock theories of arbitrary order in arbitrary spacetime dimensions involving an electromagnetic field is studied. 
General reviews on modified gravities can be found in  Refs.~\cite{Nojiri:2017ncd,Capozziello:2011et,Nojiri:2006ri}. 

Let us also call attention to Refs.~\cite{Sushkov:2009hk,Saridakis:2010mf} where modified gravities containing a nonstandard interaction between matter and gravity are under consideration. 
Depending on the choice of the parameters, such an interaction can lead to a big bang, an expanding universe with no beginning, a cosmological turnaround, 
an eternally contracting universe, a big crunch, a Big Rip avoidance, and a cosmological bounce.

In the present paper we suggest commutation relations for a quantum measure. In Sec.~\ref{quantum_measure}, we consider 
the commutation relations whose right-hand side takes account of the presence of curvature of the spacetime by introducing 
the integration of some function (depending on a metric) over the region which is the intersection of the regions
appearing on the left-hand side of the commutation relations. 
To write out these commutation relations, it is necessary to triangulate the space, as it is suggested in loop quantum gravity (LQG).
But in contrast to LQG, we, however, assume here that the space remains smooth, and the quantized measure, as well as LQG,
leads to discrete lengths, areas, and volumes. 
In Sec.~\ref{quant_meas_GR}, we study the self-consistent problem of the interaction of the quantized measure and classical gravitation. 
In Sec.~\ref{matter_quantm_measure}, we find out how the field equations change in the background of a space endowed with quantum measure.

\section{Quantization of the measure}
\label{quantum_measure}

To begin with, recall some results obtained in 
Ref.~\cite{Dzhunushaliev:2022bfk} where  the following procedure of quantization of the measure is suggested:
\begin{itemize}
	\item The triangulation of spacetime is introduced, as in LQG: the spacetime is divided into four-simplices. Each four-simplex contains tetrahedrons, triangles, and segments. 
	\item The measure $\mu$ is introduced, which in general is not associated with a metric. 
	\item The operator of this measure $\hat{\mu}$ is introduced, 
	\begin{equation}
		\widehat{\mu (\mathcal{A})} = \int \limits_{\mathcal{A}} \widehat{d \mu} ,
	\label{measure_quant}
	\end{equation}
	for which the commutation relations are imposed and where the region $\mathcal{A}$ can be tetrahedrons, triangles, and segments belonging to a four-simplex. 
\end{itemize}
Note that here, in contrast to LQG, the triangulation of spacetime is introduced as an auxiliary procedure that enables one to quantize such a nonlocal quantity like a measure. 
The spacetime by itself remains a smooth manifold which, according to the above statement, is triangulated.

Here, we would like to consider generalized commutation relations for the operator of the measure~\eqref{measure_quant}, 
\begin{equation}
	\left[
	\widehat{\mu\left( {\mathcal{A}_1}\right) }, \widehat{\mu\left( {\mathcal{A}_2}\right)}
	\right] = i l_{\text{Pl}}^\alpha 
	\widehat{
		\mu \left( \mathcal{A}_1 \cap \mathcal{A}_2 \right)
	} . 
\label{comm_rels_meas}
\end{equation}
Here the exponent $\alpha$ equalizes the dimensions of the left- and right-hand sides of the commutation relation~\eqref{comm_rels_meas}, and $l_{\text{Pl}}$ is the Planck length.
Such a definition enables one to take account of a possibility when the regions $\mathcal{A}_1$ and $\mathcal{A}_2$ may have different dimensions. 
This quantization suggests that the operators $\widehat{\mu\left( {\mathcal{A}_1}\right) }$ and  $\widehat{\mu\left( {\mathcal{A}_2}\right)}$ commutate
if the regions $\mathcal{A}_1$ and $\mathcal{A}_2$ corresponding to them do not coincide. 
If the dimensions coincide, then, by choosing as regions the lengths,  areas, and volumes of the corresponding borders of simplexes on which the space is triangulated,
we get the results known in LQG: the quantization of lengths,  areas, and volumes
(see, e.g., the textbooks~\cite{Rovelli:2014ssa,Gambini:2011zz}). 

The quantization \eqref{comm_rels_meas} does not depend on a point of spacetime on which some metric can be given;
this will lead to the fact that the quantized measure (in particular, like lengths, areas, and volumes in LQG) will not depend on the fact where 
the region $\mathcal{A}$ is located. As a quite deduced generalization of the commutation relation~\eqref{comm_rels_meas}, 
one can consider an idea that the commutation relations for the quantum measure must depend on the location of the region $\mathcal{A}$ in space: 
\begin{equation}
	\left[
	\widehat{\mu\left( {\mathcal{A}_1}\right) }, \widehat{\mu\left( {\mathcal{A}_2}\right)}
	\right] = i \gamma_0 l_{\text{Pl}}^\alpha 
	\int \limits_{ \mathcal{A}_1 \cap \mathcal{A}_2	} 
	f_g \; \widehat{d \mu} .
\label{comm_rels_meas_gen}
\end{equation}
Here $f_g$ is a function depending on the metric $g$ and $\gamma_0$ is some dimensional constant. The characteristic size of triangulation cells  is determined by the right-hand side
of the commutation relations~\eqref{comm_rels_meas_gen}, and this size should not be less than the corresponding Planck values. This implies that
 $
	\left\langle \gamma_0 
		\int \limits_{ \mathcal{A}_1 \cap \mathcal{A}_2	} 
		f_g \; \widehat{d \mu}
	\right\rangle 
	\gtrsim l_{\text{Pl}}^\beta , 
$
where $\beta$ equalizes the dimensions of the right- and left-hand sides of this inequality. In order for quantum gravity to smooth out of singularities, it would be also logical to require that
$
	f_g \rightarrow \infty 
$
as $R \rightarrow \infty$. Since the metric $g$ is a tensor and the function $f_g$ is a scalar, this function will depend on scalar invariants of the metric, such as, for example, 
$R, R_{\mu \nu}^2, R_{\mu \nu \rho \sigma}^2, \cdots$, although more complicated variants are also possible.

Such a quantization results in interesting consequences. It is logical to choose the function $f_g$ in such a way that, when the scalar invariants tend to infinity, the least possible lengths,
areas, and volumes would increase. In such case, when they tend to infinity, for instance when 
$R \rightarrow \infty$, $l_{\text{min}}, S_{\text{min}}$, and $V_{\text{min}}$ would also increase such that these singularities are smoothed out.
This means that our physical system becomes a self-consistent one in the sense that the quantized measure and metric affect each other.

\section{Self-consistent problem for a quantum measure and classical gravity
}
\label{quant_meas_GR}

Keeping in mind that the measure is quantized, the gravitational action has the following form:
\begin{equation}
	S_{g} = - \frac{1}{16 \pi G} \int R \left\langle \widehat{d \mu} \right\rangle ,
\label{GR_lagr}
\end{equation}
where $\left\langle \widehat{d \mu} \right\rangle $ is a quantum averaged measure, on account of the commutation relations~\eqref{comm_rels_meas} 
or~\eqref{comm_rels_meas_gen}, and $R$ is the scalar curvature for the metric $g$. According to the Radon-Nikodym theorem~\eqref{Radon_Nikodym}, 
the measures
$\left\langle \widehat{d \mu} \right\rangle $ and $d \mu_g = \sqrt{-g} \, d^4x$ are related by
\begin{equation}
	\left\langle \widehat{d \mu} \right\rangle  = \mathcal{F}(g) d \mu_g .
\label{RN_connection}
\end{equation}
The function $\mathcal{F}(g)$ is called the Radon-Nikodym derivative, and it depends on the metric $g$. 
Taking this relation into account, the gravitational action takes the form
\begin{equation}
	S_{g} = - \frac{1}{16 \pi G} \int R \mathcal{F}(g) d \mu_g = 
	- \frac{1}{16 \pi G} \int F(g) d \mu_g .
\label{GR_lagr_mod}
\end{equation}
Here the function $F(g)$ is a Lagrangian of modified gravity. Thus
\textit{the self-consistent problem for a quantum measure and general relativity leads to modified gravity.} 
Analogous to the arguments for the function $f_g$ from Eq.~\eqref{comm_rels_meas_gen}, we can say that the function 
$F(g)$ will depend on scalar invariants of the metric, such as, for example,
$R, R_{\mu \nu}^2, R_{\mu \nu \rho \sigma}^2, \cdots$, although more complicated variants are possible as well. This means that
$
	F(g) = F \left( R, R_{\mu \nu}^2, R_{\mu \nu \rho \sigma}^2, \cdots\right) 
$. 
An explicit form of this function and its dependence on the scalar invariants will depend on the particular physical situation: 
this can be usual modified gravity $F(R)$, or Gauss-Bonnet gravity, or Weyl gravity, etc. This important observation leads to the conclusion that 
\textit{a fixed version of modified gravity is not a fundamental theory of gravitation but it is some approximation for a description of a situation in some particular physical contexts; }
that is, for some conditions, this, for example, can be  $F(R)$ modified gravity, for other conditions~-- Weyl gravity, etc.

The function  $\mathcal{F}$ must be dimensionless, and this results in the fact that the corresponding expression contains new dimensional constants.
With the idea in hand that modified gravities are some approximations to quantum gravity, the appearance of the new dimensional
constants in modified gravities is very similar to the effect of  dimensional transmutation in quantum field theory~\cite{Coleman:1973jx}.
The relationship between nonperturbative quantization and the effect of dimensional transmutation was discussed in Ref.~\cite{Dzhunushaliev:2022apb}. 

Consistent with physical arguments, we may suppose that close to a singularity 
the function $\mathcal{F} \gg 1$.
In turn,  for pure general relativity, we must have $\mathcal{F} = 1$.

Thus, the self-consistent problem of the interaction of the quantum measure and classical gravitation is the pair: the commutation relations 
\eqref{comm_rels_meas} or \eqref{comm_rels_meas_gen} and field equations coming from the action~\eqref{GR_lagr_mod}. 

To conclude this section, we may mention the following possibility of a self-consistent interaction of the quantum measure and gravitation.
In the commutation relations~\eqref{comm_rels_meas_gen}, one can choose the function  $f_g$ to be dependent on the Einstein action,
\begin{equation}
	\left[
		\widehat{\mu\left( {\mathcal{A}_1}\right) }, \widehat{\mu\left( {\mathcal{A}_2}\right)}
	\right] = i \gamma_{0} l_{\text{Pl}}^\alpha l_0^2 
	\int \limits_{ \mathcal{A}_1 \cap \mathcal{A}_2	} 
 R \widehat{d \mu} ,
\label{comm_rels_meas_gen_GR}
\end{equation}
where $l_0$ is some dimensional constant possibly different from $l_{\text{Pl}}$. For example, $l_0^{-2}$ 
can be the average value of the curvature $\left\langle R\right\rangle $ in the region
$\mathcal{A}_1 \cap \mathcal{A}_2	$.

Then the  self-consistent problem of the interaction of the quantum measure and gravitation can be formulated as follows: 
determine the operator algebra, given by the commutation relations~\eqref{comm_rels_meas_gen_GR}, in such a way that 
the right-hand side of these commutation relations would be extremal in the following sense:
\begin{equation}
	\delta S = \delta \left\langle 
		\int \limits_{ \Omega} 
		R \; \widehat{d \mu} 
	\right\rangle = 0 , 
\label{least_action}
\end{equation}
where $\left\langle \ldots \right\rangle $ denotes the averaging over a quantum state describing some quantum measure and classical metric
satisfying the principle of least action~\eqref{least_action}. 

In the presence of matter in~\eqref{least_action}, it is necessary to add the corresponding Lagrangian~$L_m$:
$$
	\delta S = \delta \left\langle 
		\int \limits_{ \Omega} 
		\left( - \frac{R}{16 \pi G} + L_m \right) \widehat{d \mu} 
	\right\rangle = 0 . 
$$

On account of Eqs.~\eqref{RN_connection} and \eqref{GR_lagr_mod} and the arguments given after Eq.~\eqref{GR_lagr_mod}, 
the principle of least action~\eqref{least_action} can be rewritten in the form
$$
	\delta S = \delta \left\langle 
		\int \limits_{ \Omega	} 
		F(g) \; \widehat{d \mu} 
	\right\rangle 
	= \delta 
	\int \limits_{ \Omega	} 
	F(g)  \left\langle \widehat{d \mu} 
	\right\rangle 
	= 0 . 
$$

Let us now give the simplest solutions to this self-consistent problem with the commutation relations~\eqref{comm_rels_meas_gen_GR} for spaces of constant curvature
 $R = \text{const}$.

\subsection{$R = 0$. }

In this case the commutation relations~\eqref{comm_rels_meas_gen_GR} have the form
$$
	\left[
	\widehat{\mu\left( {\mathcal{A}_1}\right) }, \widehat{\mu\left( {\mathcal{A}_2}\right)}
	\right] = 0 . 
$$
This means that in flat space the measure is classical and the field equations~\eqref{mod_Euler_Lagrange} 
(which are under consideration in Sec.~\ref{matter_quantm_measure}) for matter in the background of a quantized measure
do not change.
 
\subsection{$R = \text{const} \neq 0$. }

In this case the commutation relations \eqref{comm_rels_meas_gen_GR} take the form
$$
	\left[
		\widehat{\mu\left( {\mathcal{A}_1}\right) }, \widehat{\mu\left( {\mathcal{A}_2}\right)}
		\right] = i \gamma_{0} l_{\text{Pl}}^\alpha l_0^2 R \, 
	\widehat{
		\mu \left( \mathcal{A}_1 \cap \mathcal{A}_2 \right)
	} ,
$$
where $R = \text{const}$. These are the commutation relations known in LQG; this means that on spaces of constant curvature all results
of LQG are reproduced.

\section{Matter in the background of a quantized measure
}
\label{matter_quantm_measure}

Consider now the question of how the field equations change in the background of a space endowed with quantized measure.
In analogy to~\eqref{GR_lagr} and \eqref{GR_lagr_mod}, the action for matter takes the form
\begin{equation}
	S_m =  \int L_m \left\langle \widehat{d \mu} \right\rangle 
	= \int L_m \mathcal{F}(g) d \mu_g 
	= \int L_m \mathcal{F}(R, R_{\mu \nu}^2, \cdots) d \mu_g 
	= \int L_{m, \text{mod}  } d \mu_g  . 
\label{matter_mod}
\end{equation}
Variation of the action~\eqref{matter_mod} with respect to $\phi^A$ leads to the following modified Euler-Lagrange equations:
\begin{equation}
	\frac{\partial}{\partial x^\mu} \frac{\partial L_{m, \text{mod}  }}{\partial \phi^A_{, \mu}} 
	- \frac{\partial L_{m, \text{mod}  }}{\partial \phi^A} = 0 , 
\label{mod_Euler_Lagrange}
\end{equation}
where $\phi^A$ are matter fields with a generalized index $A$. It is evident that the presence of the quantized measure in the form of the factor
 $\mathcal{F}$ modifies the Euler-Lagrange field equations. According to the discussion given in Sec.~\ref{quant_meas_GR}, 
 in high curvature regions, the Euler-Lagrange equations differ considerably from the similar equations in flat space.
 
Modified theories with the action \eqref{matter_mod} have been considered in Refs.~\cite{Sushkov:2009hk,Saridakis:2010mf} using the Lagrangian
$$
	L = \frac{R}{8 \pi} - F_{\mu \nu} \phi^{, \mu} \phi^{, \nu} ,
$$
where the function $F_{\mu \nu}$ describes a nonminimal derivative coupling between the scalar field~$\phi$ and gravitation. 
In those papers, two cases are under consideration: 
$
	F_{\mu \nu} = g_{\mu \nu} + \kappa_1 g_{\mu \nu} R 
	+ \kappa_2 R_{\mu \nu}
$ and 
$
F_{\mu \nu} =	\epsilon g_{\mu \nu} + \kappa G_{\mu \nu}
$, where $G_{\mu \nu}$ is the Einstein tensor. The results of calculations within these theories indicate that such interactions enable one, for example,
to avoid a cosmological singularity and a Big Rip. 

Note also that in the flat space limit $R \rightarrow 0$ the right-hand side of the commutation relations~\eqref{comm_rels_meas_gen_GR} 
becomes equal to zero, and the measure  $\mu$ becomes classical; this enables one to choose the function $\mathcal{F}(g) = 1$ in Eq.~\eqref{RN_connection}, 
and then \eqref{matter_mod} becomes the action for usual unmodified matter.

\section{Conclusions}

We have considered the quantization of the measure and suggested the commutation relations for operators of the measure. From the physical point of view,
it is quite natural to assume that a measure and a metric must somehow affect each other. To realize this idea, we have considered the 
self-consistent problem of the influence of the quantum measure and classical gravity on each other and shown in this context that such a self-consistent problem leads
to modified theories of gravitation. It is important that \textit{in the regions where gravitational forces differ strongly the modified gravities will be different. }

Along the lines suggested here, we have also considered the influence of the quantum measure on the Euler-Lagrange equations for matter fields. 
It became clear that in the regions with strong gravitational fields these equations  differ considerably from similar equations in flat space. 

Thus, in the present paper, we have  
\begin{itemize}
	\item suggested the commutation relations for the operator algebra of measure;
	\item considered the self-consistent problem of the interaction of the quantized measure and classical gravitation;
	\item considered one of versions of the commutation relations whose right-hand side contains the action of general relativity; 
	\item solved the self-consistent problem of the interaction of the quantized measure and classical gravitation in the simplest case of spaces of constant curvature;
	\item shown that the quantized measure leads to modified gravities;
	\item considered the problem of field equations for an arbitrary matter field in the background of a space endowed with quantum measure.
\end{itemize}

In the context of the present study, there appear numerous problems, requiring further investigations: (a)~Is it possible, using the technique developed in LQG,
to get spectra of lengths, areas, and volumes in a space endowed with a metric and with the use of the  commutation relations~\eqref{comm_rels_meas_gen}
which take account of this metric? (b)~Are the commutation relations~\eqref{comm_rels_meas} and \eqref{comm_rels_meas_gen} 
mathematically consistent in the case when the regions $\mathcal{A}_1$ and $\mathcal{A}_2$ have different dimensions?
 (c)~Do the commutation relations~\eqref{comm_rels_meas} and \eqref{comm_rels_meas_gen} lead to smoothing out of singularities?
 (d)~What is the relationship between the functions $f_g$ and $\mathcal{F}$? etc. 

Another interesting question arising in this context concerns a possible relationship between Weyl gravity and an approximate description of a self-consistent 
interaction of the quantum measure and classical gravitation near singularities. The reason is that Weyl gravity is conformally invariant and 
it may not feel the appearance of singularities, as shown in Ref.~\cite{Dzhunushaliev:2021cgu}; 
this can be related to smoothing out of the singularities while quantizing a measure.

The next interesting question: are there maximum lengths, areas, and volumes arising while quantizing a measure?
In LQG, the answer to this question is negative, but we know that in quantum mechanics situations with finite spectrum of some physical quantity are possible;
therefore, a priori,  one cannot rule out such a situation while quantizing a measure interacting with gravitation (metric).

\section*{Acknowledgments}

The work was supported by the Science Committee of the Ministry of Science and Higher Education of the Republic of Kazakhstan (Grant  No.~AP14869140, ``The study of QCD effects in non-QCD theories''). 

\appendix

\section{Radon-Nikodym theorem}	
Let us introduce the definition of a measure:
\begin{definition}
	Let $X$ be a set and $\Sigma$ a $\sigma$-algebra over $X$. A function $\mu$ from $\Sigma$ to the extended real number line is called a measure if it satisfies the following properties:
	\begin{itemize}
		\item Non-negativity: For all $E$ in $\Sigma$, we have $\mu(E) \geq 0$.
		\item Null empty set: $ \mu (\varnothing )=0$.
		\item Countable additivity (or $\sigma$-additivity): For all countable collections
		$\left\lbrace E_{k}\right\rbrace_{k=1}^{\infty }$ of pairwise disjoint sets in $\Sigma$,
			$$			\mu \left(
			\bigcup_{k=1}^{\infty } E_{k}
			\right) = \sum_{k = 1}^{\infty } \mu (E_{k}) .$$
	\end{itemize}
\end{definition}
The pair $\left( X, \Sigma\right) $ is called a measurable space. A triple $\left( X, \Sigma, \mu\right) $ is called a measure space.
The Radon-Nikodym theorem involves a measurable space $(X,\Sigma )$ on which two $\sigma$-finite measures are defined, $\mu$ and $\nu$.
\begin{theorem}[the Radon-Nikodym theorem~\cite{Rudin}]
	If $\nu \ll \mu$ (that is, if $\nu$  is absolutely continuous with respect to~$\mu$), then there exists a $\Sigma$-measurable function $f:X\to [0,\infty )$,
	such that for any measurable set $\mathcal A \subseteq X$,
	\begin{equation}
		\nu (\mathcal A) = \int\limits_{\mathcal A} f\,d\mu .
	\end{equation}
	\label{Radon_Nikodym}
\end{theorem}
The function $f$ appearing here is called the Radon-Nikodym derivative and is designated as $\frac{d \nu}{d \mu}$.
This theorem implies that any two measures can be interrelated by some function.


\begin{thebibliography}{99}
	
\bibitem{Starobinsky:1980te}
A.~A.~Starobinsky,
A New Type of Isotropic Cosmological Models Without Singularity,
Phys. Lett. B \textbf{91}, 99 (1980). 

\bibitem{Nojiri:2017ncd}
S.~Nojiri, S.~D.~Odintsov and V.~K.~Oikonomou,
Modified Gravity Theories on a Nutshell: Inflation, Bounce and Late-time Evolution,
Phys. Rept. \textbf{692}, 1 (2017).

\bibitem{Guendelman:1996qy}
E.~I.~Guendelman and A.~B.~Kaganovich,
The Principle of nongravitating vacuum energy and some of its consequences,
Phys. Rev. D \textbf{53}, 7020 (1996).

\bibitem{Guendelman:1996jr}
E.~I.~Guendelman and A.~B.~Kaganovich,
Gravitational theory without the cosmological constant problem,
Phys. Rev. D \textbf{55}, 5970 (1997).

\bibitem{Guendelman:2012gg}
E.~Guendelman, D.~Singleton, and N.~Yongram,
A two measure model of dark energy and dark matter,
JCAP \textbf{11}, 044 (2012).

\bibitem{Finster:2023rkv}
F.~Finster, E.~Guendelman and C.~F.~Paganini,
Modified Measures as an Effective Theory for Causal Fermion Systems,
[arXiv:2303.16566 [gr-qc]].

\bibitem{Gronwald:1997ei}
F.~Gronwald, U.~Muench, A.~Macias and F.~W.~Hehl,
Volume elements of space-time and a quartet of scalar fields,
Phys. Rev. D \textbf{58}, 084021 (1998). 

\bibitem{Nojiri:2003ft}
S.~Nojiri and S.~D.~Odintsov,
Modified gravity with negative and positive powers of the curvature: Unification of the inflation and of the cosmic acceleration,
Phys. Rev. D \textbf{68}, 123512 (2003). 

\bibitem{Lidsey:2002zw}
J.~E.~Lidsey, S.~Nojiri and S.~D.~Odintsov,
Brane world cosmology in (anti)-de Sitter Einstein-Gauss-Bonnet-Maxwell gravity,
JHEP \textbf{06}, 026 (2002). 

\bibitem{Chakraborty:2020ifg}
S.~Chakraborty and N.~Dadhich,
Limits on stellar structures in Lovelock theories of gravity,
Phys. Dark Univ. \textbf{30}, 100658 (2020). 

\bibitem{Capozziello:2011et}
S.~Capozziello and M.~De Laurentis,
``Extended Theories of Gravity,''
Phys. Rept. \textbf{509}, 167 (2011). 

\bibitem{Nojiri:2006ri}
S.~Nojiri and S.~D.~Odintsov,
Introduction to modified gravity and gravitational alternative for dark energy,
eConf \textbf{C0602061}, 06 (2006). 

\bibitem{Sushkov:2009hk}
S.~V.~Sushkov,
Exact cosmological solutions with nonminimal derivative coupling,
Phys. Rev. D \textbf{80}, 103505 (2009). 

\bibitem{Saridakis:2010mf}
E.~N.~Saridakis and S.~V.~Sushkov,
Quintessence and phantom cosmology with non-minimal derivative coupling,
Phys. Rev. D \textbf{81}, 083510 (2010). 

\bibitem{Dzhunushaliev:2022bfk}
V.~Dzhunushaliev and V.~Folomeev,
Quantization of Measure in Gravitation,
Grav. Cosmol. \textbf{29},  367 (2023).

\bibitem{Rovelli:2014ssa}
C.~Rovelli and F.~Vidotto,
\textit{Covariant Loop Quantum Gravity: An Elementary Introduction to Quantum Gravity and Spinfoam Theory}
(Cambridge University Press, 2015).

\bibitem{Gambini:2011zz}
R.~Gambini and J.~Pullin,
\textit{A first course in loop quantum gravity}
(Oxford University Press, 2011).

\bibitem{Coleman:1973jx}
S.~R.~Coleman and E.~J.~Weinberg,
Radiative Corrections as the Origin of Spontaneous Symmetry Breaking,
Phys. Rev. D \textbf{7}, 1888 (1973). 

\bibitem{Dzhunushaliev:2022apb}
V.~Dzhunushaliev and V.~Folomeev,
QCD Effects in Non-QCD Theories,
Found. Phys. \textbf{52}, 118 (2022). 

\bibitem{Dzhunushaliev:2021cgu}
V.~Dzhunushaliev and V.~Folomeev,
Masking singularities in Weyl gravity and Ricci flows,
Eur. Phys. J. C \textbf{81}, 387 (2021).  

\bibitem{Rudin}
W. Rudin,
\textit{Real and Complex Analysis}
(New York: McGraw-Hill, pp. 116-134, 1986).
	
\end{thebibliography}
\end{document}